Verification of $\Gamma_7$ symmetry assignment for the top valence band of ZnO by magneto-optical studies of the free *A* exciton state

Lu Ding,[1,a] Chunlei Yang,[1,b] Hongtao He,[1,c] Fengyi Jiang,[2] Jiannong Wang,[1] Zikang Tang,[1] Bradley A. Foreman,[1] and Weikun Ge[1,d,*]

[1]Department of Physics, Hong Kong University of Science and Technology, Clear Water Bay, Kowloon, Hong Kong, People's Republic of China

[2]Institute of Materials Science, Nanchang University, Nanchang 330047, Jiangxi, People's Republic of China

The circularly-polarized and angular-resolved magneto-photoluminescence spectroscopy was carried out to study the free *A* exciton 1*S* state in wurtzite ZnO at 5 K. Fine spectral lines due to linear Zeeman splitting were resolved with ultrahigh resolution. By measuring the circular polarization of Zeeman splitting lines, the top valence band is verified to have $\Gamma_7$ symmetry. The out-of-plane component $B_\parallel$ of the magnetic field, which is parallel to ZnO's *c* axis, leads to linear Zeeman splitting of both the dipole-allowed $\Gamma_5$ exciton state and the weakly allowed $\Gamma_1/\Gamma_2$ exciton states. The in-plane field $B_\perp$, which is perpendicular to the *c* axis, increases the oscillator strength of the weak $\Gamma_1/\Gamma_2$ states by forming a mixed exciton state.

I. INTRODUCTION

Zinc oxide continues to attract strong interests for optoelectronic applications thanks to its direct and wide-bandgap nature and large exciton binding energy. Its electronic and optical properties have been studied for many years, with a sharp increase in activity during the past decade.[1,2] Despite its long history, some fundamental properties of ZnO are still not fully understood. The valence-band symmetry ordering is especially controversial. In most wurtzite semiconductors, the quasidegenerate *p*-like valence states at $\Gamma$ are split by the crystal-field ($\Delta_{cr}$) and spin-orbit ($\Delta_{so}$) interactions into states of symmetry $\Gamma_9$, $\Gamma_7$, and $\Gamma_7$, in order of decreasing energy[3]. For ZnO, it is widely accepted that $\Delta_{cr} \gg |\Delta_{so}|$, but the sign of $\Delta_{so}$ has been the subject of much debate. Thomas[4] and Hopfield[5], in their pioneering reflectivity studies of excitonic transitions, proposed that ZnO has a negative spin-orbit splitting, leading to a reversed $\Gamma_7 - \Gamma_9 - \Gamma_7$ ordering.

This reversed ordering is consistent with a wide variety of experimental data (see Refs. 6-13 for a few examples), and is also supported by first-principles calculations[14]. Nevertheless, some authors[15-22] have rejected this interpretation, even in the most recent published work[23], in favor of the conventional $\Gamma_9 - \Gamma_7 - \Gamma_7$ ordering (*i.e.* $\Delta_{so} > 0$). Many of the studies supporting reversed ordering did not directly compare the two possibilities; hence, although these studies provide cumulative evidence in favor of reversed ordering, they cannot be said to definitively resolve the controversy. Some such studies also used models with a large number of fitting parameters, leaving open the possibility that other parameter sets (perhaps consistent with a different ordering) might yield an equally good fit. A further reason for the continued skepticism toward models with $\Delta_{so} < 0$ may be that they often depend crucially on parameters (*e.g.*, *k*-linear terms[6,7,11,12]) that are very small in the limit $\Delta_{cr} \gg |\Delta_{so}|$.

A more direct approach was taken in Refs. 24 and 25, which used first-principles calculations[24] and magneto-optical studies of bound excitons[25] (BX) to argue that the sign of the hole *g* factor deduced from magneto-optical studies of free excitons (FX) in Ref. 16 is incorrect, and that the top valence band of ZnO should therefore have $\Gamma_7$ symmetry. Recently, the result reported by Ref. 13 on a newly observed hole state related Zeeman fine splitting for BXs in the Voigt configuration together with angular- and polarization-dependent magneto-optical measurements also provides strong evidence for the $\Gamma_7$ symmetry of the top valence band. However, as pointed out by Thomas and Hopfield[26], the hole $g$ factors derived from studies of BX may, in principle, be entirely different from the $g$ factors of free holes, due to mixing of the quasidegenerate valence states by the defect potential. For this reason, it is not *a priori* obvious that results based on BX are capable of providing unambiguous evidence for the symmetry of the top valence band of ZnO.

In view of the simple and well defined nature of FX, we have employed high-resolution magneto-photoluminescence (PL) of free *A* excitons to investigate the valence-band ordering in a more straightforward manner. A powerful technique, magneto-PL explicitly reveals the relationship between the fundamental optical transitions of semiconductors and the optical selection rules that are uniquely determined by the band structure symmetries. In this paper, we present evidence obtained from careful and detailed magneto-PL measurements that strongly supports the conclusion that the top valence band of wurtzite ZnO has $\Gamma_7$ symmetry. This interpretation is also supported by the polarization dependence of the Zeeman splitting of neutral-impurity BX.[30] A brief description of the results given here was previously posted on an online archive.[27]

II MAGNETO-PHOTOLUMINESCENCE SPECTROSCOPY

Free excitons involving the *s*-like $\Gamma_7$ conduction band and the three valence bands are labeled as *A*, *B*, and *C* excitons, in order of increasing exciton energy[4]. Depending on the symmetry assigned to the top valence band, the *A* excitons have two possible symmetries:

$$\Gamma_7 \otimes \Gamma_7 \rightarrow \Gamma_5 \oplus \Gamma_1 \oplus \Gamma_2, \; \Gamma_7 \otimes \Gamma_9 \rightarrow \Gamma_5 \oplus \Gamma_6. \quad (1)$$

Here the doubly degenerate $\Gamma_5$ exciton is dipole-allowed for light polarized normal to the hexagonal $c$ axis ($\vec{E} \perp \vec{c}$) and the singly degenerate $\Gamma_1$ exciton is (weakly) dipole-allowed for $\vec{E} \parallel \vec{c}$, whereas the doubly degenerate $\Gamma_6$ exciton and the singly degenerate $\Gamma_2$ exciton are dipole-forbidden.

Using a magneto-cryostat with magnetic field $B$ up to 7 T, the magneto-PL measurements were performed on a 3 $\mu$m thick high quality ZnO thin film deposited on (0001) sapphire substrate using metal-organic chemical vapor deposition (MOCVD). The inset of Fig. 1(a) depicts the magneto-PL experimental setup. The Faraday configuration ($\vec{k} \parallel \vec{B}$) is applied, where $\vec{k}$ is the wave vector of the emitted light and $\theta$ is the angle between $\vec{B}$ and the $c$ axis. $\vec{B}$ can be decomposed into an out-of-plane component $B_\parallel = B \cos\theta$ (parallel to the $c$ axis) and an in-plane component $B_\perp = B \sin\theta$ (perpendicular to the $c$ axis). In our setup, different angles $\theta$ were achieved by simply rotating the $c$ axis. The incident laser was perpendicular to the magnetic field for arbitrary $\theta$, except that the backscattering geometry was used for $\theta = 0$. The magneto-PL spectra were resolved by a monochromator (SPEX 1403) located along the $\vec{B}$ field direction with 1800 g/mm double gratings and detected by a photomultiplier tube (R928). The spectral resolution of the system is about 0.1 meV. The circular polarization ($\sigma_+$ or $\sigma_-$) of the emitted light was analyzed using a quarter-wave plate and a linear polarizer. All the measurements were performed at 5 K to minimize energy shifts induced by thermal fluctuation.

III. RESULTS AND DISCUSSIONS

To demonstrate clearly the magnetic field effect, the angular-dependent zero-field PL as well as magneto-PL spectra of the $A$ exciton 1S state ($FX_A^{n=1}$) are shown for comparison in Figs.

1(a) and 1(b), respectively. At $B = 0$ T, two resolved fine structures of $FX_A^{n=1}$ are labeled as $P_1$ (3.3757 eV, weak) and $P_2$ (3.3778 eV, strong), which correspond to the weakly allowed (or dipole-forbidden) and dipole-active excitons, respectively [see Fig. 1(a)]. The changes of the peak positions and intensities are found to be negligible at different $\theta$, which indicates a weak dependence on the polarization direction of the incident laser. Applying a magnetic field of 7 T, rich features are found with strong angular dependence in the PL spectra [see Fig. 1(b)]. When $\theta = 10°$, Zeeman splitting of $P_1$ is observed with a splitting energy $\Delta E_{P_1}$ as large as 1.4 meV, whereas $P_2$ remains nearly unchanged. When $\theta$ increases, $\Delta E_{P_1}$ becomes smaller. The two split peaks of $P_1$ finally merge into one at $\theta = 80°$. On the other hand, the integrated intensity $I_{P_1}$ of $P_1$ increases with increasing $\theta$ and eventually dominates the $FX_A^{n=1}$ spectrum. It is worth noting that there is almost no change in the magneto-PL spectrum at $\theta = 0°$ when $B$ is scanned from 0 T to 7 T, which is due to the weakly allowed (or dipole-forbidden) nature of $P_1$ at $B_\perp = 0$. The in-plane magnetic field $B_\perp$ is found to significantly increase the oscillator strength of $P_1$.

Our experimental data are best explained by a perturbation model with (1) $\Gamma_7 - \Gamma_9 - \Gamma_7$ valence-band ordering (*i.e.* the spin-orbit splitting $\Delta_{so} < 0$) and (2) a weak magnetic field case. In this model, the uncontroversial assumptions are made that the crystal-field splitting $\Delta_{cr}$ is much greater than the spin-orbit splitting $\Delta_{so}$, which in turn is much greater than either the exchange splitting or the Zeeman splitting (for the magnetic fields used in our experiment). All of these perturbations are assumed to be negligible in comparison to the fundamental energy gap of the crystal.

We treat $\Delta_{so}$ as a perturbation of $\Delta_{cr}$, working to first order in the energy and to zeroth order in the state vector. If we choose the $z$ and $c$ axes to be the same, the exciton states formed from the $p_x \pm i p_y$ hole states of $\Gamma_7$ symmetry (i.e., the A excitons according to Thomas[4] and Hopfield[5]) are therefore

$$|\Gamma_5^{(7)}, \pm\rangle = |\pm\rangle|\pm 1, \mp\rangle \; (g_{exc} = g_h^\parallel + g_e), \quad (2a)$$

$$|\Gamma_{1\oplus 2}, \pm\rangle = |\mp\rangle|\pm 1, \mp\rangle \; (g_{exc} = g_h^\parallel - g_e). \quad (2b)$$

Here $|+\rangle|m_h, -\rangle$ is the tensor product of a spin-up $s$ electron and a spin-down $p$ hole whose $z$ component of orbital angular momentum is $m_h$. The $\pm$ label of the exciton states is taken from the sign of $m_h$ (note that for $\Gamma_5$, $m_h$ is also the $z$ component of the total exciton angular momentum). In Eq. (2b), the contribution of $\Delta_{so}$ to the short-range exchange interaction is neglected, so that $\Gamma_1$ and $\Gamma_2$ form an approximately doubly-degenerate reducible representation[4,5,23,24] denoted $\Gamma_{1\oplus 2}$. A small field $B_\parallel$ produces a linear Zeeman splitting with the given exciton effective $g$ factors $g_{exc}$, in which $g_e$ is the (nearly) isotropic electron $g$ factor and $g_h^\parallel$ is the hole $g$ factor parallel to the $c$ axis[25]. In the simple model of Ref. 24 we have $g_h^\parallel = 2K - g_0$, where $K = -(3\kappa + 1)$ is the magnetic Luttinger parameter and $g_0 = 2$ is the $g$ factor of a free hole. The states in Eq. (2b) are dipole-forbidden when $B_\perp = 0$, but they become dipole-allowed $B_\perp \neq 0$ due to mixing with $|\Gamma_5^{(7)}, \pm\rangle$ caused by $g_e$.

Likewise, the exciton states formed from the $p_x \pm i p_y$ hole states of $\Gamma_9$ symmetry (i.e., the B excitons according to Thomas and Hopfield) are given by

$$|\Gamma_5^{(9)}, \pm\rangle = |\mp\rangle|\pm 1, \pm\rangle \; (g_{exc} = g_h^\parallel - g_e), \quad (3a)$$

$$|\Gamma_6, \pm\rangle = |\pm\rangle|\pm 1, \pm\rangle \; (g_{exc} = g_h^\parallel + g_e). \quad (3b)$$

in which $g_h^\parallel = 2K + g_0$. Just as for $|\Gamma_{1\oplus 2}, \pm\rangle$, the states $|\Gamma_6, \pm\rangle$ are dipole-forbidden when $B_\perp = 0$, but become dipole-allowed when $B_\perp \neq 0$ due to $g_e$-induced mixing with $|\Gamma_5^{(9)}, \pm\rangle$.

In Fig. 2, we sketch two sets of optically allowed exciton transitions in a magnetic field with arbitrary $\theta$ (so that $B_\parallel$ and $B_\perp$ are both nonzero) for the ground-state free excitons involving a hole of either (a) $\Gamma_7$ symmetry or (b) $\Gamma_9$ symmetry. Here $\delta$ is defined as the zero-field exchange splitting between $\Gamma_5$ and $\Gamma_{1\oplus 2}$ states in case (a) or between $\Gamma_5$ and $\Gamma_6$ states in case (b). The labels $\pm 1/2$ and $\pm 3/2$ in Fig. 2 refer to the z component of total angular momentum for conduction and valence electrons. The n notation $\sigma_\pm^*$ indicates that these transitions are dipole-forbidden when $B_\perp = 0$, but emit photons with $\sigma_\pm$ polarization when $B_\perp \neq 0$. The sign of $g_h^\parallel$ would have to be negative for $\Gamma_7$ and positive for $\Gamma_9$ in order to agree with the experimental observation that the Zeeman splitting of the weakly allowed states is much larger than that of the dipole-active states.

Based on the information in Eqs. (2) and (3) and the energy diagrams in Figs. 2(a) and 2(b), it is evident that the symmetry of the top valence band can be identified by measuring the polarization of the weakly allowed under an applied magnetic field. For exciton transitions involving a $\Gamma_9$ hole and a $\Gamma_7$ electron, one would expect the originally dipole-forbidden states ($\Gamma_6$ excitons) to split, with the lower-energy peak showing $\sigma_-$ polarization. However, if both the electron and hole have $\Gamma_7$ symmetry, the originally weakly allowed $\Gamma_{1\oplus 2}$ excitons will show $\sigma_+$ polarization for the lower-energy peak.

Figure 2(c) presents the polarization dependence of the magneto-PL of the $A$ exciton state with $B = 3$ T and $\theta = 45°$. The two dominant Zeeman-split peaks of $P_1$ are well resolve in this figure. This clearly indicates that the lower-energy peak of $P_1$ has $\sigma_+$ polarization, which unambiguously demonstrates that the hole in the $A$ exciton 1S state (or the top valence band) in wurtzite ZnO has $\Gamma_7$ symmetry. $P_1$ is therefore attributed to $\Gamma_{1\oplus 2}$, where as $P_2$ is attributed to $\Gamma_5$.

The experimentally determined zero-field exchange splitting $\delta$ is 2.1 meV [see Fig. 1(a)], which is in good agreement with Refs. 2, 4, and 28.

To get more information on the electron and hole $g$ factors, the magnetic field dependences of the transition energies of $P_1$ and $P_2$ are summarized in Figs. 3(a) and 3(c) for $\theta = 20°$ and $\theta = 80°$, respectively. Figure 3(b) shows the $\theta$ dependence of $P_1$ and $P_2$ at $B = 7$ T. In the Zeeman splitting of $P_1$ and $P_2$, $B_\parallel$ lifts the degeneracy of the $P_1$ ($\Gamma_{1\oplus 2}$) states or the doublet $P_2$ ($\Gamma_5$) state. The energy splitting of $P_1$ ($\Gamma_{1\oplus 2}$) is fitted using $E_{P_{1\pm}} = E_{P_1} \pm \frac{1}{2}(g_h^\parallel - g_e)\mu_B B_\parallel$, where $\mu_B$ is the Bohr magneton and $E_{P_1} = 3.37576$ eV is the zero-field transition energy of $P_1$ ($\Gamma_{1\oplus 2}$). Using $g_e = 1.95$[29], the hole $g$ factor obtained from the fitting (see solid curves in Fig. 3) is $g_h^\parallel = -1.6$, which agrees well with the values obtained in Refs. 6 and 7 (but with a different convention for the sign of $g_h^\parallel$). The fact that the Zeeman splitting for $P_2$ ($\Gamma_5$) could not be resolved (see open circles in Fig. 3) indicates the nearly equal absolute values of $g_e$ and $g_h^\parallel$. The dotted curves for $P_2$ are plotted according to $E_{P_{2\pm}} = E_{P_2} \pm \frac{1}{2}(g_h^\parallel + g_e)\mu_B B_\parallel$, employing $g_e = 1.95$ and $g_h^\parallel = -1.6$.

In addition, the Zeeman splitting of BXs $I_6$ and $I_7$ has also been observed and the transition energies are shown in Fig. 3. The circular polarization dependences indicate that $I_6$ and $I_7$ are excitons bound to neutral impurity centers with $A$ holes involved[30]. The dashed lines are fitted results given by $\pm \frac{1}{2}\mu_B B(g_e + g_h)$ and $g_h = g_h^\parallel \sqrt{\cos^2\theta + (g_h^\perp/g_h^\parallel)^2 \sin^2\theta}$, where $g_e = 1.95$, $g_h^\parallel = -1.6$, and $g_h^\perp = 0.11$. The equality of the fitted FX and BX values of $g_h^\parallel$ provides *ex post facto* support for the conclusions of Ref. 25 (although, as noted in the introduction, such similarity cannot be assumed to hold in general).

The different contributions of the in-plane and out-of-plane magnetic field to the magneto-PL spectra are shown more specifically in Fig. 4. The left panel [Fig. 4(a)] shows the measured dependence of the Zeeman splitting $\Delta E_{P_1}$ of $P_1$ ($\Gamma_{1\oplus 2}$). The data taken at $B = 7$ T for different $\theta$ (solid black dots) and those taken at fixed $\theta$ for different $B$ (hollow colored dots) fall onto the same line plotted using the equation $\Delta E_{P_1} = |g_{exc}\mu_B B_\parallel|$ with $g_{exc} = g_h^\parallel - g_e = -3.55$. The zero-field splitting of the $\Gamma_1$ and $\Gamma_2$ states is zero as expected. This good linear relationship between $\Delta E_{P_1}$ and $B_\parallel$ indicates that the weak-magnetic-field condition is well satisfied and reveals that the splitting of the $A$ exciton states depends on the out-of-plane field instead of the total magnetic field, which can be well explained by the $\Gamma_2$ symmetry of the out-of-plane field that mixes $\Gamma_1$ only with $\Gamma_2$ states[5]. Figure 4(b) shows that the intensity $I_{P_1}$ of $P_1$ increases monotonically with increasing $B_\perp$. The transition probability of the originally weakly allowed $\Gamma_1/\Gamma_2$ excitons increases significantly due to mixing with $\Gamma_5$ excitons.

## IV. CONCLUSIONS

We applied the magneto-PL spectroscopy on a high quality ZnO thin film and carried out the circular polarization and angular-resolved analysis. The top valence band of wurtzite ZnO was verified to have $\Gamma_7$ symmetry by directly examining the polarization of the free $A$ exciton emission. The out-of-plane component $B_\parallel$ of the magnetic field was found to be responsible for the linear Zeeman splitting of the $\Gamma_5$ and $\Gamma_1/\Gamma_2$ states. The in-plane magnetic field $B_\perp$ increases the oscillator strength of the originally weakly allowed $\Gamma_1/\Gamma_2$ states by mixing with $\Gamma_5$ states. The hole effective $g$ factor was found to be negative and has the value $-1.6$.

ACKNOWLEDGMENTS

The authors are grateful to Y. Zhang (The University of North Carolina at Charlotte, USA) and G. Q. Hai (Universidade de São Paulo, Brazil) as well as Y. Q. Wang (Institute of Solid State Physics, Chinese Academy of Sciences, China) for encouraging discussions. This work is funded by the Hong Kong University of Science and Technology via grants No. DAG04/05.SC24 and DAG05/06.SC30.

FIG. 1: (Color online) Angular-dependent PL spectra of $FX_A^{n=1}$ at (a) $B = 0$ T and (b) $B = 7$ T, taken at $T = 5$ K. Inset of (a) shows the experimental setup. All spectra are normalized at the higher-energy side of $FX_A^{n=1}$.

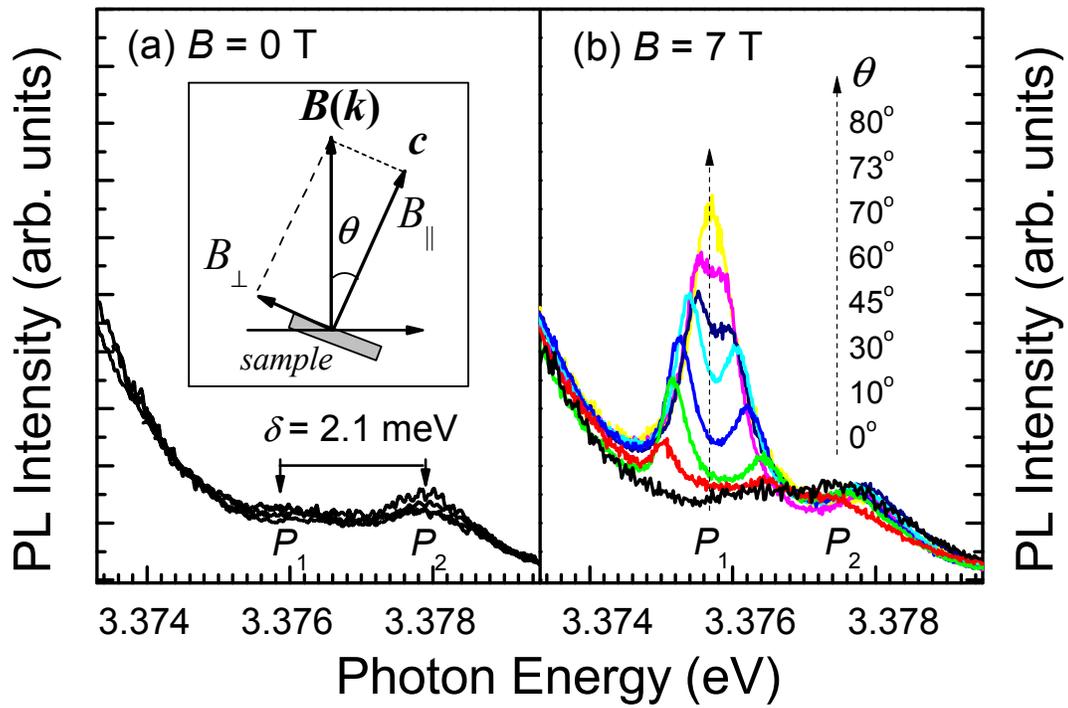

FIG. 2: (Color online) Schematic representations of energy levels of $A$ exciton transitions involving holes of (a) $\Gamma_7$ symmetry and (b) $\Gamma_9$ symmetry. (c) shows the circular polarization dependence of the magneto-PL of $FX_A^{n=1}$.

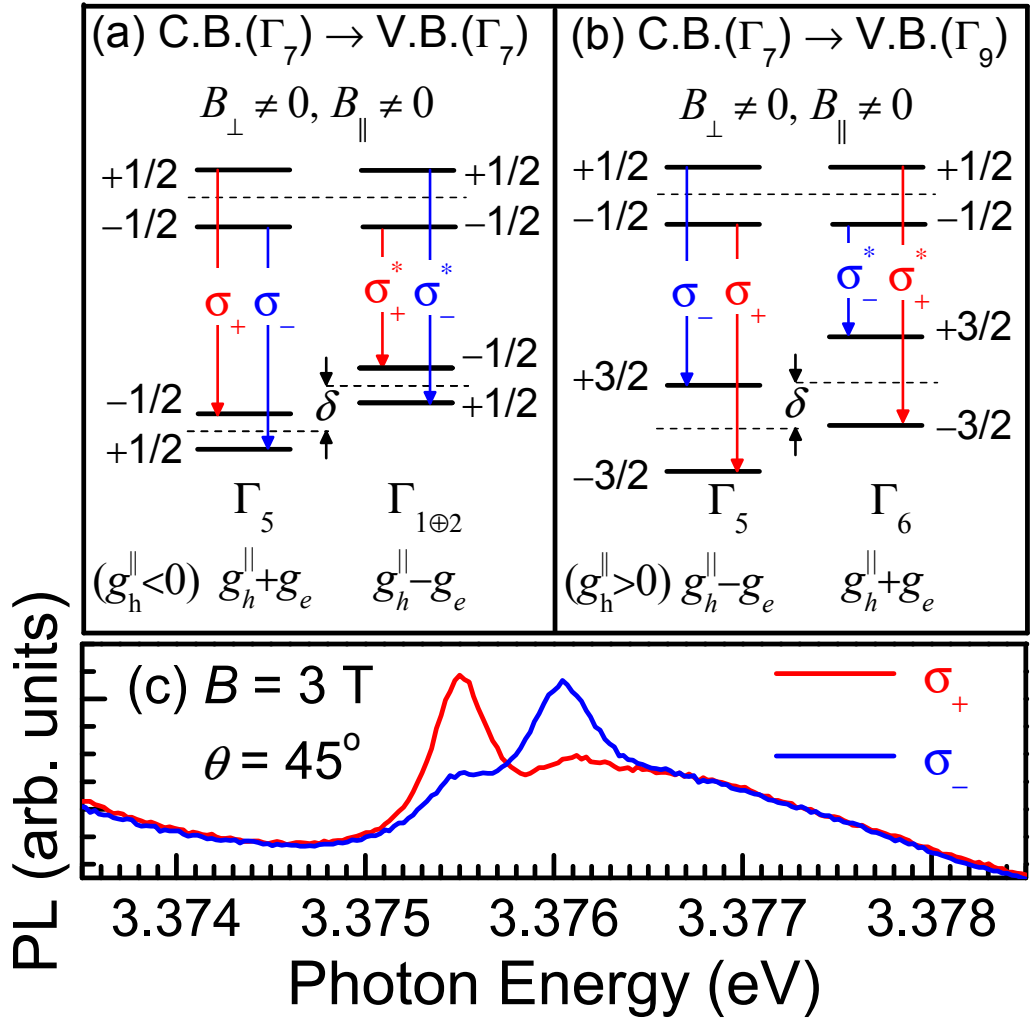

FIG. 3: (Color online) The magnetic field and angular dependences of the peak energies of $A$ excitons ($P_1$ and $P_2$) and BXs ($I_6$ and $I_7$), as described in the text.

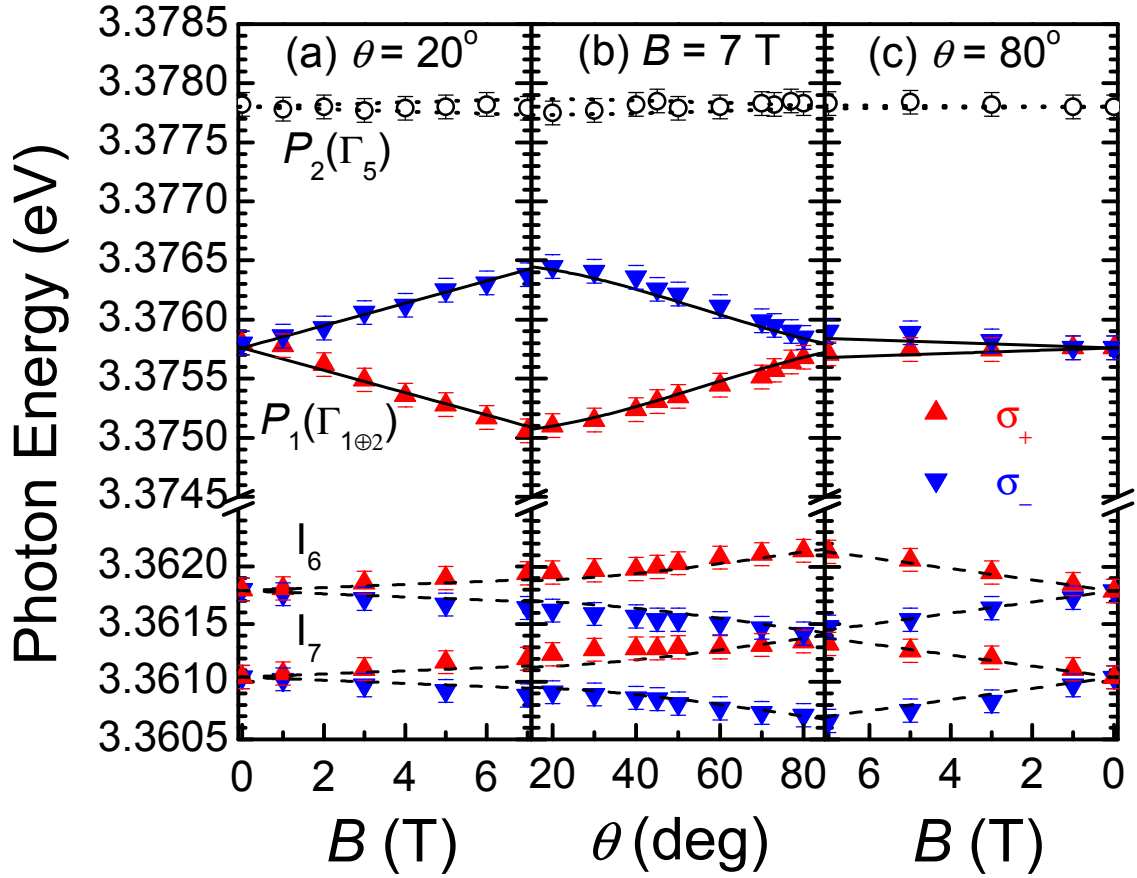

FIG. 4: (Color online) (a) $B_\parallel$ dependence of the Zeeman splitting $\Delta E_{P_1}$ of $P_1$ ($\Gamma_{1\oplus 2}$). (b) $B_\perp$ dependence of the intensity $I_{P_1}$ of $P_1$ ($\Gamma_{1\oplus 2}$) (solid dots). The dashed line is a guide for the eyes.

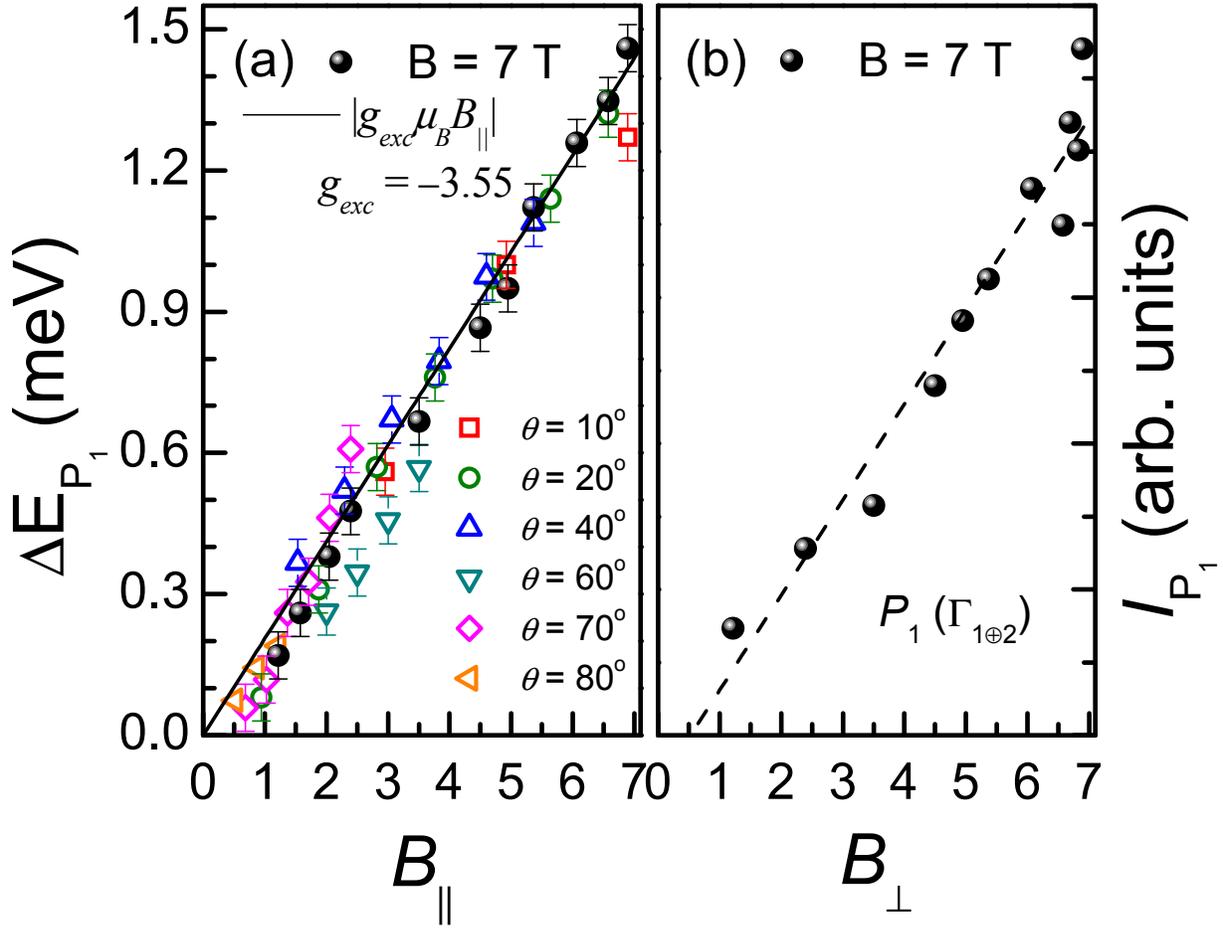

[a]Present address: Institute of Materials Research and Engineering, Agency for Science, Technology and Research (A*STAR), Singapore 117602

[b]Present address: Center for photovoltaic solar cells, Shen Zhen Institute of Advanced Technology, Chinese Academy of Sciences, Shenzhen 518055, China

[c]Present address: Department of Physics, The South University of Science and Technology of China, Shenzhen, China

[d]Present address: Department of Physics, Tsinghua University, Beijing 100871, China

[*]phweikun@ust.hk